\documentstyle[epsf,11pt,aas2pp4]{article}

\received{July 24, 1996}
\accepted{November 1, 1996}

\lefthead{M\'endez \& van der Klis}
\righthead{The EXOSAT data on GX~339--4}

\begin{document}

\title{The EXOSAT data on GX~339--4: further evidence for an
       ``intermediate'' state} 

\author{Mariano M\'endez\altaffilmark{1,2}
        \and
        Michiel van der Klis\altaffilmark{1}
}

\altaffiltext{1}{Astronomical Institute ``Anton Pannekoek'',
       University of Amsterdam and Center for High-Energy Astrophysics,
       Kruislaan 403, NL-1098 SJ Amsterdam, the Netherlands}

\altaffiltext{2}{Facultad de Ciencias Astron\'omicas y Geof\'{\i}sicas, 
       Universidad Nacional de La Plata, Paseo del Bosque S/N, 
       1900 La Plata, Argentina}

\authoraddr{Dr. Mariano M\'endez
            Astronomical Institute ``Anton Pannekoek''
            University of Amsterdam
            Kruislaan 403
            NL-1098 SJ Amsterdam
            The Netherlands
            Phone : +31 20 525 7475         (office phone)
                    +31 20 525 7491/7492    (Secretary's office phone)
            Fax   : +31 20 525 7484
            E-mail: mariano@astro.uva.nl
            Dr. Michiel van der Klis
            Astronomical Institute ``Anton Pannekoek''
            University of Amsterdam
            Kruislaan 403
            NL-1098 SJ Amsterdam
            The Netherlands
            Phone : +31 20 525 7498         (office phone)
                    +31 20 525 7491/7492    (Secretary's office phone)
            Fax   : +31 20 525 7484
            E-mail: michiel@astro.uva.nl
}

\begin{abstract}

We have studied the fast timing and spectral behavior of the black
hole candidate (BHC) GX~339--4 using all 1--20\,keV {\sl EXOSAT\/} ME
data:  the July 1983, March 1984, May 1984 and April 1985
observations.  In April 1985 GX~339--4 was in a weak low state (0.03
10$^{-9}$ erg cm$^{-2}$s$^{-1}$, 2--10 keV).  The X-ray spectrum was a
hard power law with a photon index $\alpha$ of 1.82 and the power
spectrum, though ill-constrained by the data, was consistent with a
typical BHC low state spectrum.  During the other three pointings the
system was brighter (1.5 10$^{-9}$ erg cm$^{-2}$s$^{-1}$, 2--10 keV),
an ultrasoft component was present in the spectrum, and the power law
was steeper ($\alpha\sim$3.5 in 1984; in 1983 it was not
well-measured); the power spectrum showed a flat-topped band-limited
noise component with a break frequency of $\sim$4.5\,Hz and a
fractional rms amplitude of $\sim$7\%.  Comparing flux levels, X-ray
spectra and power spectra we conclude that in these observations
GX~339--4 was in a state intermediate between the usual BHC low and
high states.  The system was $\sim 4$ times brighter than in its usual
low state, 4 times fainter than in its usual high state, and an order
of magnitude fainter than in its very high state.  We compare these
results to those recently obtained on other BHCs, and conclude that
this ``intermediate state'' behavior is a common characteristic of
BHCs, that occurs at $\dot M$ levels intermediate between the high and
the low state.  We argue that this result can be used to resolve the
long-standing issue of the dependence of the power spectral break
frequency in the low state on mass accretion rate, and strengthens the
idea that low-state noise and very-high state noise may have a common
origin.  We briefly discuss a possible interpretation for the changes
in break frequency in the low state and between low state and
intermediate state.

\end{abstract}


\section{Introduction}

Discovered with the OSO--7 satellite (Markert et al.
\cite{markert73}), GX~339--4 is considered a black hole candidate
(BHC) due to the characteristics of its fast X-ray variability, and
high-low (soft-hard) state transitions similar to Cyg~X--1.  Its X-ray
behavior has been studied with various satellites (Markert et al.
\cite{markert73}, Samimi et al.  \cite{samimi79}, Nolan et al.
\cite{nolan82}, Motch et al.  \cite{motch83}, Ricketts
\cite{ricketts83}, Maejima et al.  \cite{maejima84}, Makishima et al.
\cite{makishima86}, Ilovaisky et al.  \cite{ilovaisky86}, Miyamoto et
al.  \cite{miyamoto91} and \cite{miyamoto92}).  The source has shown
large random variability on time-scales of milliseconds up to months
(Markert et al.  \cite{markert73}, Samimi et al.  \cite{samimi79},
Motch, Ilovaisky, \& Chevalier \cite{motch82}, Motch et al.
\cite{motch83}), and quasi-periodic oscillations (QPOs) have been
reported with frequencies of $\sim 6.25$ Hz (Makishima et al.
\cite{makishima88}), $\sim 0.8$ Hz (Grebenev et al.
\cite{grebenev91}), $\sim 0.1$ Hz (Motch et al.  \cite{motch83}, Motch
et al.  \cite{motch85}, Imamura et al.  \cite{imamura90}) and $\sim
0.05$ Hz (Motch et al.  \cite{motch83}).  The optical counterpart of
GX~339--4, a star of V $\sim 18$ mag.  discovered by Doxsey et al.
(\cite{doxsey79}), was also searched for fast variability and QPOs
(e.g., Motch et al.  \cite{motch83}, Motch et al.  \cite{motch85}),
and in some cases the results were similar to X-ray data obtained
simultaneously.  Optical periodicities of 1.13 ms (Imamura et al.
\cite{imamura87}) and 190 s (Steiman-Cameron et al.  \cite{steiman90})
were reported, however these results have not been confirmed.
Finally, photometric data revealed a 14.8-h modulation of the
brightness of the optical counterpart (Callanan et al.
\cite{callanan92}), which may be the orbital period of the system.

Despite the fact that the compact object mass has not been measured,
GX~339--4 is a prototype BHC in that it has shown all three of the
classical black hole ``states'' (see van der Klis \cite{vanderklis95}
for a review):  (1) the low state (LS), with a flat ($\alpha$=1.5--2)
power law X-ray spectrum (Tananbaum et al.  \cite{tananbaum72}) and
strong (25--50\% rms) band limited noise (Oda et al.  \cite{oda71})
with a break frequency of 0.03--0.3 Hz; (2) a high state (HS) where
the $2 - 10$~keV flux is an order of magnitude higher than in the LS
due to the presence of an ultrasoft X-ray spectral component with
sometimes an $\alpha$=2--3 power law tail, and a weak (few \% rms)
flat power law power spectrum; and (3) a very high state (VHS;
Miyamoto et al.  \cite{miyamoto91}) characterized by high X-ray
luminosity (2--8 times higher than in the HS), an ultrasoft X-ray
spectral component plus a ($\alpha\sim$2.5) power law tail, strongly
variable 1--15\% rms band limited noise with a much higher cut off
frequency (1--10 Hz) than in the LS, and 3--10 Hz QPO.  The 2--10 keV
flux is higher in the HS than in the LS, but at higher energies the
situation is often the reverse, and in some sources, (e.g., GX~339--4;
Grebenev et al.  \cite{grebenev93}) the integrated 1--200 keV
luminosity is higher in the LS than in the HS.  The order in which
black hole transients have been observed in their decay to go through
these states, and also similarities to neutron star states strongly
suggest, however, that the accretion rate is highest in the VHS, lower
in the HS and lowest in the LS (van der Klis \cite{vanderklis94a}).
As noted by van der Klis (\cite{vanderklis95}), there is some
ambiguity in the way the HS has been defined.  When only spectral data
and no time variability data were available, the HS was loosely
defined as any state where the ultrasoft component was not negligible
compared to the power law component in the 2--10 keV range.  When the
variability was also measured, it turned out that the HS was also
characterized by an absence of band-limited noise such as in the LS;
any band limited noise in the HS was thought to be a characteristic of
the power law X-ray spectral component and therefore stronger at
higher photon energy (Oda et al.  \cite{oda76}, Miyamoto
\cite{miyamoto94}), so this absence of band-limited noise, as
observed, for example, in the HS of GS~1124--68 was also used as a HS
criterion.  Of course, for the observations classified as HS with no
variability information, we can not be sure of the noise properties.

\begin{deluxetable}{ccccccc}
\tablecolumns{7}
\scriptsize
\tablecaption{Observations of GX~339--4 carried out with {\sl EXOSAT\/}.
\label{tabobs}}
\tablewidth{0pt}
\tablehead{
\colhead{Start} & \colhead{End} & \colhead{OBC} &
\colhead{Time resolution\tablenotemark{a}} & \colhead{Count rate} &
\colhead{Nr. of detectors} & \colhead{Detector} \\
\colhead{Date, UT} & \colhead{Date, UT} & \colhead{modes} &
\colhead{[1/1024 s]} & \colhead{[c/s]} & 
\colhead{on source} & \colhead{I.D.} \\
}
\startdata
1983 Jul\phs 17, 01:40      & 1983 Jul\phs 17, 06:44      & ~~~HER4, HTR3 &
  ~8  & 888.4 & 4 & Ar + Xe \\
1984 Mar 12, 09:47          & 1984 Mar 12, 16:10          & ~~~HER5, HTR3 &
  ~8  & 270.2 & 4 & Ar      \\
1984 May\phs 4, 04:15       & 1984 May\phs 4, 10:26       & ~~~HER5, HTR3 &
  ~8  & 296.2 & 4 & Ar      \\
1985 Apr\phd\phd 29, 11:54  & 1985 Apr\phd\phd 29, 17:56  & HER4/6, HTR3  &
  32  & 612.8 & 4 & Ar + Xe \\
\enddata
\tablenotetext{a}{HTR3 mode}
\end{deluxetable}

A characteristic property of the LS band-limited noise is its variable
break frequency; the power spectrum above the break changes very
little, whereas below the break the power spectrum is approximately
flat and power is ``missing'' from the power spectrum (Belloni \&
Hasinger \cite{belloni90a}, Miyamoto et al.  \cite{miyamoto92}).
These break frequency variations, which anticorrelate with the level
of the flat top take place without any correlation with the 2--20 keV
flux (Belloni \& Hasinger \cite{belloni90a}) or X-ray spectrum (van
der Klis \cite{vanderklis94a}), but there is a correlation with the
40--145 keV spectral properties (Crary et al.  \cite{crary96}).  On
the basis of the similarities between LS and VHS noise, and
similarities with neutron star phenomenology, van der Klis (
\cite{vanderklis94a}) suggested that the break frequency increases
with mass accretion rate.

GX~339--4 seemed to fit right in with this picture (Ilovaisky et al.
\cite{ilovaisky86}, Nolan et al.  \cite{nolan82}, Miyamoto et al.
\cite{miyamoto91}).  In what follows we analyze all {\sl EXOSAT\/} ME
data on GX~339--4 and study the relation between the X-ray spectrum
and the power spectrum.  This work was motivated by the peculiar
characteristics (high break frequency and low rms at an intermediate
X-ray flux) of the power spectrum of this source presented by Belloni
\& Hasinger (\cite{belloni90b}).  (A similar situation existed with
the 1991 May 17, power spectrum of GS~1124--68 presented by Miyamoto
et al.  \cite{miyamoto94}; see Belloni et al.  \cite{belloni96a} for a
report on the power spectral characteristics of that source).
Although some of the {\sl EXOSAT\/} data of GX~339--4 have been
reported previously (Ilovaisky et al.  \cite{ilovaisky86} discussed
the X-ray spectra of May 1984 and April 1985, while Belloni \&
Hasinger \cite{belloni90b} published the power spectra of July 1983
and March 1984), in none of these papers a full comparison of timing
and spectral data was performed.  As we will argue in Section 4, when
viewed in the light of our current understanding of black-hole
candidate phenomenology, the EXOSAT data on GX~339--4 provide new
insight into the nature of the transition between low and high state
in black hole candidates.

\section{Observations and data reduction}

\begin{deluxetable}{cccc}
\tablecolumns{4}
\tablecaption{Lorentzian fitting to the power spectra of GX~339--4.
\label{tabpow}}
\tablewidth{0pt}
\tablehead{
\colhead{Date}        &
\colhead{rms [\%]}   &
\colhead{HWHM [Hz]}      &
\colhead{Reduced $\chi^2$} \\
}
\startdata
Jul\phs 83     & $6.9 \pm 1.2$ & $5.0 \pm 2.6$ & 0.90 \\
Mar\phd 84     & $7.5 \pm 0.5$ & $3.9 \pm 0.6$ & 1.03 \\
May 84         & $6.6 \pm 0.6$ & $5.2 \pm 1.5$ & 0.95 \\
Apr\phd\phd 85 & $<$ 26        & --            & --   \\
\enddata
\end{deluxetable}

The Medium-Energy (ME) experiment on board {\sl EXOSAT\/} consisted of
an array of eight detectors (each with an argon and a xenon-filled
proportional counter) with a total area of $1600$ cm$^{2}$ that gave
moderate spectral resolution in the 1--50 keV band (Turner, Smith, \&
Zimmermann \cite{turner81}, White \& Peacock \cite{white88}).  The
experiment was divided in two halves, each consisting of four
detectors.  In our observations, one half was pointed at the source
and the other half was offset to measure the background.  Sometimes
``array swaps'' were performed where the two halves switched role.
Data were processed by an On Board Computer (OBC) which had different
modes emphasizing either spectral (HER, High Energy Resolution) or
timing (HTR, High Time Resolution) information.  GX~339--4 was
observed four times with this instrument.  A log of the observations
is given in Table~\ref{tabobs}.  The OBC programs used for the data
presented here were HTR3, HER4 and HER5.  For the spectral studies we
used the HER data from the 1--20\,keV argon detectors.  For the timing
studies we used the HTR3 data.  In this mode counts detected by both
halves, aligned and offset, were accumulated without energy
information.  HTR3 count rates sometimes include only the argon
detectors, and sometimes both argon and 5--50 keV xenon detectors
summed together (see Table~\ref{tabobs}).  The xenon detectors
contributed mainly background.

\subsection{Timing data}

\begin{deluxetable}{cccccc}
\tablecolumns{6}
\footnotesize
\tablecaption{X-ray spectra of GX~339--4.
\label{tabspe}}
\tablewidth{0pt}
\tablehead{
\colhead{Date}        &
\colhead{N$_{\rm H}$ [$10^{22}$ cm$^{-2}]$} &
\colhead{$kT$ [keV]}          &
\colhead{$n$ (photon-index)} &
\colhead{Flux\tablenotemark{a} [erg cm$^{-2}$ s$^{-1}$]}  &
\colhead{Reduced $\chi^2$} \\
}
\startdata
Mar\phd 84     & 0.78 (0.68 -- 0.89)               & 0.42 (0.41 -- 0.43) &
 3.74 (3.57 -- 3.89) & $1.51 \times 10^{-9~}$                      & 1.15 \\
May 84         & 0.55 (0.46 -- 0.65)               & 0.42 (0.41 -- 0.43) &
 3.17 (2.97 -- 3.36) & $1.50 \times 10^{-9~}$                      & 0.83 \\
Apr\phd\phd 85 & 0.51 (0.33 -- 0.82)               & --                  &
 1.82 (1.64 -- 2.02) & $3.10 \times 10^{-11}$                      & 1.11 \\
\enddata
\tablenotetext{a}{Unabsorbed total flux in the 2--10 keV range.}
\tablenotetext{}{3$\sigma$ confidence ranges are indicated in brackets.}
\end{deluxetable}

To study the timing behavior we divided the data into segments of 8192
contiguous bins, preserving the full time resolution $\Delta t$
available in each case (see Table~\ref{tabobs}).  Each segment was
checked for gaps and spikes (due to instrumental effects), where a
spike was defined as a single-bin excess over the average count rate
with a probability of occurrence from Poisson statistics of $<10^{-8}$
per bin.  Data segments with a spike or a gap (only a few percent in
each observation) were excluded from further analysis.  We produced
power spectra by Fourier transforming each segment, and squaring and
Leahy-normalizing the resulting Fourier transforms.  Our power spectra
thus span from $(8192 \Delta t)^{-1}$ to $(2 \Delta t)^{-1}$ Hz.  We
subtracted the deadtime affected Poisson noise (Kuulkers et al.
\cite{kuulkers94}) and the instrumental high-frequency noise (Berger
and van der Klis \cite{berger94}) individually from each power
spectrum, and then averaged the corrected power spectra (see van der
Klis \cite{vanderklis89}).  We finally renormalized the average power
spectra to (rms/mean)$^2$/Hz normalization (e.g., van der Klis
\cite{vanderklis95}), logarithmically rebinned them and fitted them
using power laws, Lorentzians and various combinations of these
functions.

\subsection{Spectral data}

We extracted and background subtracted the HER spectra using the {\sl
EXOSAT\/} Interactive Analysis (IA) software (Parmar, Lammers, \&
Angelini \cite{parmar95}).  For the April 1985 data the background was
determined from the same detectors as used for recording the source
spectrum, using an array swap.  We corrected for the small differences
due to detector tilt (``difference spectra''; Parmar et al.,
\cite{parmar95}).  For the March and May 1984 data no array swap was
performed, so slew data were used instead to obtain a background
estimate.  We carefully examined these for possible variations, and
excluded times when the background was unstable.  During the July 1983
observation neither method could be applied, and the spectrum obtained
was of bad quality, so it was not used in our further analysis.  All
reduced spectra were compared to those available from the archives at
the High Energy Astrophysics Science Archive Research Center (HEASARC)
in Goddard, and the agreement was excellent.

\section{Results}

\subsection{Power spectra}

\begin{figure}[t]
\plotfiddle{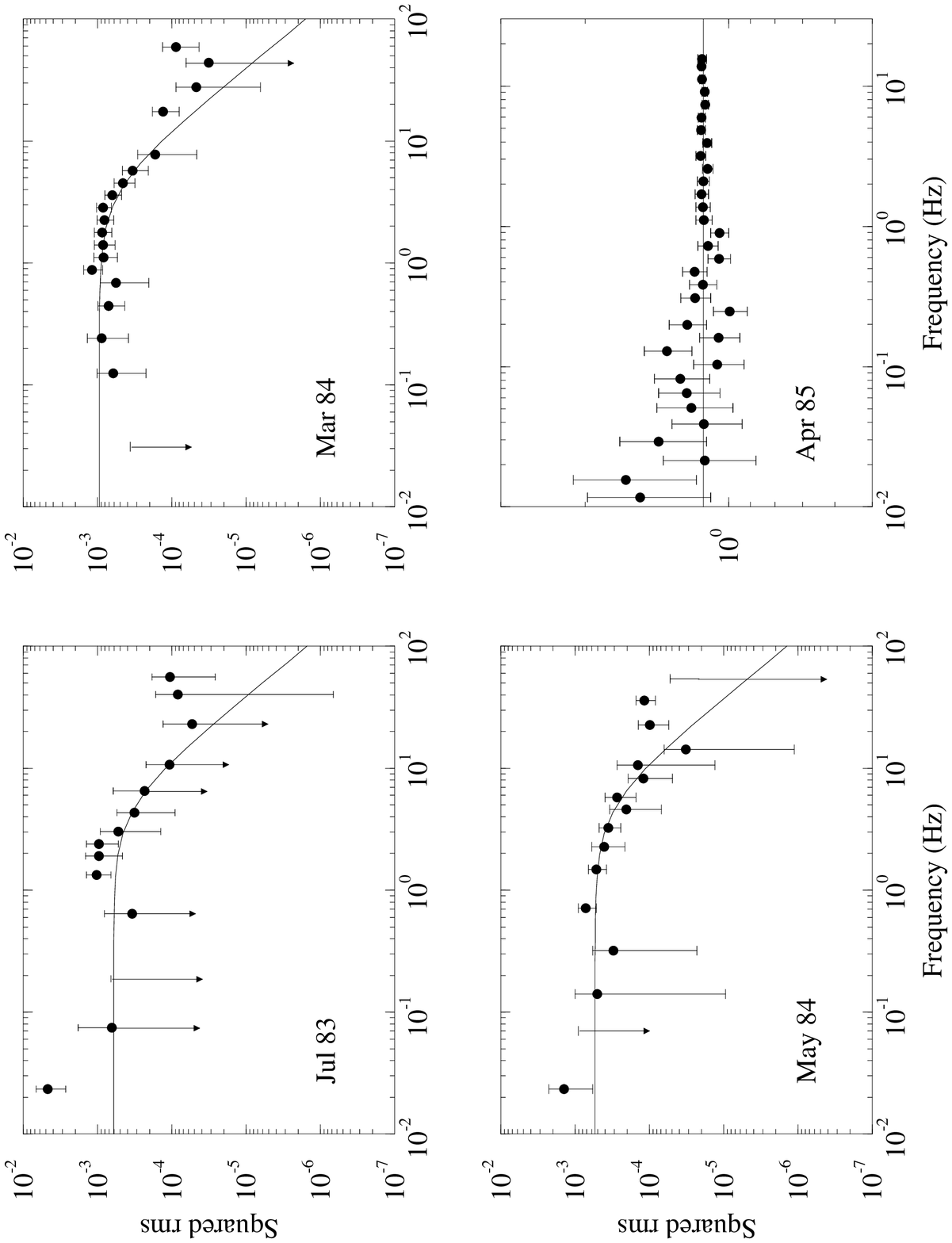}{250pt}{270}{70}{70}{-120}{400}
\caption{The power spectra of GX~339--4. (a) July 1983,
(b) March 1984, (c) May 1984, (d) April 1985.}
\label{figpow}
\end{figure}

All but the 1985 power spectra showed a band limited noise component
with an approximately flat top extending up to 2--4 Hz that gradually
steepened towards higher frequencies, with no noticeable peaks.  In
all three cases a good fit was attained using a single zero-centered
Lorentzian (a broken power law fitted as well), with a HWHM
(half-width at half maximum) of about 4.5 Hz and a root mean square
variability (rms) of about 7\% (0.001--60 Hz).  The best fit
parameters are given in Table~\ref{tabpow}.  The power spectrum of
April 1985 was ill-constrained, as the source count rate was quite
low.  The 3$\sigma$ upper limit to the 0.002--10 Hz power corresponded
to an rms amplitude of 26\%.  All power spectra are shown in
Fig.~\ref{figpow}.

\subsection{X-ray spectra}

We fitted the X-ray spectra with the sum of a blackbody and a
power law function to represent the ultra-soft and the hard component,
respectively, plus interstellar absorption.  This choice does not mean
that we argue in favor of such a model over physically motivated
models.  However, the model provides a good fit to our data and its
simplicity allows us to easily analyze the relative importance of each
component as a function of source state.

As the ME experiment was not sensitive to low absorption columns, we
made a simultaneous fit to the ME and Low-Energy (LE) data available
for the 1984 and 1985 observations.  In 1984 LE data were taken using
the 3000 lexan, the aluminium/parylene and the boron filters, each of
them with different spectral responses.  In 1985 only the 3000 lexan
filter was used.

\begin{figure}[t]
\plotfiddle{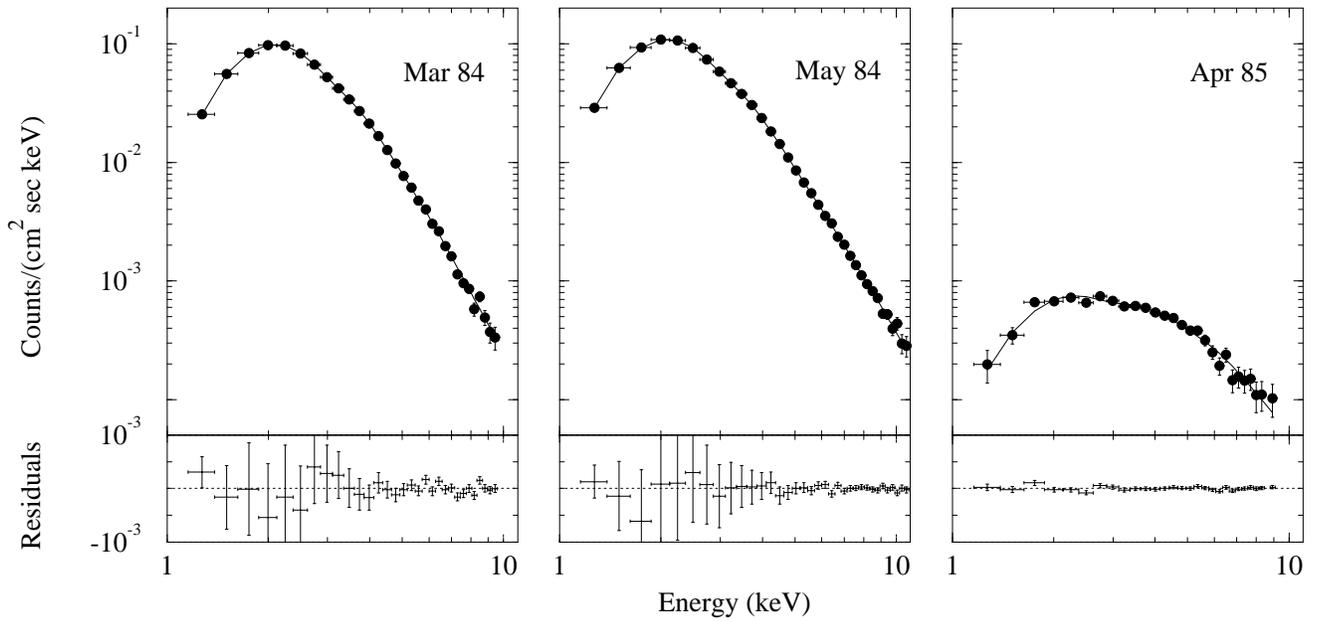}{250pt}{270}{70}{70}{-140}{400}
\caption{The X-ray spectra of GX~339--4. (a) March 1984,
                        (b) May 1984, (c) April 1984.}
\label{figspe}
\end{figure}

The best fit parameters are given in Table~\ref{tabspe}, and the
spectra are shown in Fig.~\ref{figspe}.  The slopes of the fitted
power laws of the March and May 1984 spectra are significantly
different, but both much steeper ($\alpha\sim$3.5) than that of April
85 ($\sim$1.8).  In April 1985 no soft component is detected (this was
already noted by Ilovaisky et al.  \cite{ilovaisky86}).

Almost in all cases the absorption in the line of sight is in
accordance, within the quoted error bars, to the value of $N_{\rm H} =
(5.0 \pm 0.7)~10^{21}$ cm$^{-2}$ obtained by Ilovaisky et al.
(\cite{ilovaisky86}).

\section{Discussion}
\label{secdiscussion}

\begin{deluxetable}{lccccccccccccc}
\tablecolumns{6}
\scriptsize
\tablecaption{Parameters for the four different states in BHC's.
\label{tabgen}}
\tablewidth{0pt}
\tablehead{
\colhead{}                                    &
\multicolumn{4}{c}{GX~339--4}                  &
\colhead{}                                    &
\multicolumn{3}{c}{Cyg~X--1 \tablenotemark{a}} &
\colhead{}                                    &
\multicolumn{4}{c}{GS~1124--68}                \\
\cline{3-4}
\cline{7-9}
\cline{12-13} \\
\colhead{}    &
\colhead{LS}  &
\colhead{IS}  &
\colhead{HS}  &
\colhead{VHS} &
\colhead{}    &
\colhead{LS}  &
\colhead{IS}  &
\colhead{HS \tablenotemark{b}}  &
\colhead{}    &
\colhead{LS}  &
\colhead{IS}  &
\colhead{HS}  &
\colhead{VHS} \\
}
\startdata
Flux ($2-10$ keV) \tablenotemark{c} &
  1 &
  4 &
 19 &
 54 &
\colhead{} &
  1 &
  3 &
  5 &
\colhead{} &
  1 &
  6 &
 35 &
248 \\
n (photon index) & 
1.6 &
3.5 &
2.0 &
2.5 &
\colhead{} &
1.5 &
2.2 &
2.7 &
\colhead{} &
1.6 &
2.3 &
2.4 &
2.3 \\
rms [\%] & 
18 &
 7 &
$<2$ &
10 &
\colhead{} &
30 &
20 &
\nodata &
\colhead{} &
20 &
 6 &
$<1$ &
 8 \\
$\nu_{{\rm cut}}$ [Hz] &
$<1$ &
 5 & 
0 \tablenotemark{d} &
2 &
\colhead{} &
$<1$ &
5 &
\nodata &
\colhead{} &
0.2 &
 3 &
 0 \tablenotemark{d} &
 2 \\

\tablecomments{All values quoted are approximate values.}
\tablenotetext{a}{Cyg~X--1 was never observed in the VHS.}
\tablenotetext{b}{No published power spectrum of Cyg~X--1 in the HS.
In view of the lack of timing data, the 2--10 keV flux suggests that
previously reported Cyg~X--1 HS observations may also be interpreted
as IS.}
\tablenotetext{c}{In units of the flux in the LS. The LS fluxes of
GX~339--4, Cyg~X--1 and GS~1124--68 are 0.4, 7.1 and 0.8 
$10^{-9}$ erg~cm$^{-2}$~sec$^{-1}$ respectively.}
\tablenotetext{d}{Power spectrum is a power law.}
\tablerefs{
GX~339--4: Maejima et al. \cite{maejima84}; Makishima et al.
\cite{makishima86}; Grebenev et al. \cite{grebenev91};
Miyamoto et al. \cite{miyamoto91}; Iga et al. \cite{iga92};
this paper;
Cyg~X--1: Ogawara et al. \cite{ogawara82}; Belloni et al.
\cite{belloni96b};
GS~1124--68: Miyamoto et al. \cite{miyamoto94}.}
\enddata

\end{deluxetable}

Our analysis indicates that in April 1985 GX~339--4 was probably in a
weak LS.  Although this was usually called `off state' because the
2--10 keV flux is $\sim$10 times lower than in the LS, the ratio of
the X-ray to optical luminosity (Motch et al.  \cite{motch85}), the
energy spectrum (compare Table \ref{tabspe} of this paper to Table 1
in Iga, Miyamoto, \& Kitamoto \cite{iga92}) and the total rms (upper
limit, Table \ref{tabpow}) are consistent with that of a typical BHC
low state.  More sensitive timing data would be necessary to determine
if this is in fact a different state, or just a LS at a lower mass
accretion rate.  However, the July 1983, March and May 1984
observations do not fit in with any of the classical BHC states
discussed in Section 1.  The presence of an ultrasoft X-ray spectral
component, the steepness of the power law X-ray spectral component and
the weakness and high break frequency of the band-limited noise make
these observations very different from the LS.  The presence of the
band limited noise shows that the source was not in the HS.  The X-ray
spectral and power spectral characteristics seem to be most similar to
a VHS, but the total 2--10 keV flux is an order of magnitude lower
than that in the VHS (Miyamoto \cite{miyamoto91}), and in fact {\it in
between} previously reported LS (Iga et al.  \cite{iga92}) and HS
(Makishima et al.  \cite{makishima86}) levels, a factor $\sim$4 away
from each.

Although it was not possible to analyze the energy spectrum of the
July 1983 data, the flux level and the characteristics of the power
spectrum were similar to the 1984 observations, suggesting that then
the system was in a similar state.

GX~339--4 in these observations seems to have been in a similar state
to that which GS~1124--68 showed on 1991 May 17, with a flux
intermediate between the LS and the HS, an X-ray spectrum consisting
of an ultrasoft component and a power law tail (Ebisawa et al.
\cite{ebisawa94}), and a power spectrum similar to that of the VHS
with a break frequency of a few Hz (Miyamoto et al.
\cite{miyamoto94}).  A 6~\% rms QPO at 6.7 Hz was recently discovered
in GS~1124--68 in this state (Belloni et al.  \cite{belloni96a}).  The
lower count rates would have prevented us from detecting a similar
feature in GX~339--4 if it had been present.  As its X-ray flux
decayed, GS~1124--68 subsequently went through the VHS, HS, and LS
(Miyamoto et al.  \cite{miyamoto93}), demonstrating the connection
between source state and mass accretion rate; the May 17 observation
also in time occurred in between the HS and the LS.

Based on this comparison we conclude that during July 1983 and March
and May 1984 GX~339--4 was also in a state intermediate between the LS
and HS.  We will refer to this state as Intermediate State (IS).  It
is characterized by a 2--10 keV flux intermediate between LS and HS, a
two component energy spectrum (consisting of a soft component and a
power law), and a band-limited power spectrum that is flat up to a few
Hz, and then decays at higher frequencies.  In Table \ref{tabgen} we
compare the general properties of the different states in GX~339--4,
and in two other BHC's, Cyg~X--1 and GS~1124--68.  From this Table it
can be seen that in all cases the IS flux is in between that in the LS
and that in the HS (and thus much lower than in the VHS).  The IS
cut-off frequency is always much higher (and the rms lower) than in
the LS, and in all three cases the energy spectrum is softer in the IS
than in the LS.  However we note that the power law slope in GX~339--4
(3.5) is different from that in Cyg~X--1 and GS~1124--68 (2.2--2.3).
Due to the narrow spectral coverage, the actual value of the power law
slope in GX~339--4 depended upon the model adopted for the soft
component, ranging from $\sim 2.2$ for an unsaturated Comptonized
spectrum, to $\sim 3.5$ for a blackbody spectrum.  Nevertheless it
must be stressed that a power law was always needed, and that no good
fit to the data was possible by means of a single component model.
Except for the flux, which is $\sim10 - 40$ times lower, the
timing and spectral parameters in the IS are similar to those in the
VHS.  Although this may lead to interpret the IS as a weak VHS, this
explanation should also account for the existence of a HS in between.

On the basis of a comparison of the anticorrelation between break
frequency and power density at the break observed in Cyg~X-1 (Belloni
\& Hasinger \cite{belloni90a}) and the values of these two parameters
in the VHS of GX~339--4 and GS~1124--68, van der Klis
(\cite{vanderklis94a}) proposed that the noise in these two states
might follow the same relation of break-frequency vs.  power-density
at the break, and are perhaps due to the same physical process.  The
intermediate state observations in GX~339--4 and GS~1124--68 fit in
with the VHS results (Fig~\ref{b&h}), and by their closer association
with the LS (closer in flux in both sources, and also in time sequence
in GS~1124--68), strengthen this idea.  The implication of this of
course is, that there exists a correlation with mass accretion rate,
where the break frequency increases and the rms decreases as the
accretion rate increases (see also van der Klis \cite{vanderklis94a}),
which would resolve the long-standing issue of the dependence of the
break frequency on mass accretion rate in the LS.  The problem with
that interpretation is the absence of systematic 1--20 keV X-ray
spectral variations in the LS which would indicate a systematic change
in accretion rate with break frequency.  However, the recent results
of Crary et al.  (\cite{crary96}) show that there {\it is} a
correlation with the 45--140 keV X-ray spectral properties, a steeper
power law index and lower flux correlating with a higher break
frequency.  We note, that extrapolating the Crary et al.  results to
lower energy, the spectra are seen to pivot around a point near 40 keV
(van der Klis \cite{vanderklis96}), so that on that basis one would
expect a positive correlation between break frequency and 2--10 keV
count rate.  Of course, this is not observed, perhaps due to the
presence of a weak variable ultrasoft component even in the LS.  Our
proposed interpretation of a positive correlation between break
frequency and accretion rate in the LS and IS would imply that the
45--140 keV flux drops and its slope steepens when the accretion rate
increases.  This is similar to what is already known to be the case in
the LS to HS transitions.

Finally, we consider the relation between break frequency vs.  power
at the break among all black hole candidates.  The power spectra of
many black hole candidates show various bumps and wiggles, and even
clear QPO peaks (e.g., van der Hooft et al.  \cite{vanderhooft96}) at
frequencies around the break frequency.  We point out that in spite of
these complications, it is still possible to define the width, or the
break frequency of such power spectra, for example by extrapolating
the high and low frequency parts of the power spectrum to where they
intersect (the method we used), or by fitting a broad smooth shape and
ignoring the residual peaks.  Using this approach, we have collected
the relevant data on all black hole candidate power spectra in the LS,
IS and VHS, and plotted them in Fig~\ref{b&h}.  Clearly, there is a
trend that is common among all black hole candidates towards higher
break frequency corresponding to lower power density at the break.

We point out, that in some models (e.g., Narayan \cite{narayan96}),
the inner radius of the accretion disk is expected to move out as the
accretion rate decreases, with radii as large as thousands of km (an
advection dominated flow region is supposed to form inside this
radius), so that if the break frequency is identified with the
Keplerian frequency at this radius, this would be in accordance with
the observed correlation.  The 40--145 keV X-ray spectrum becomes
steeper and the flux in that range weaker as the break frequency
increases; it would be of great interest to see what such models
predict in this respect.

\section{Conclusion}
\label{secconclusion}

\begin{figure}[t]
\plotfiddle{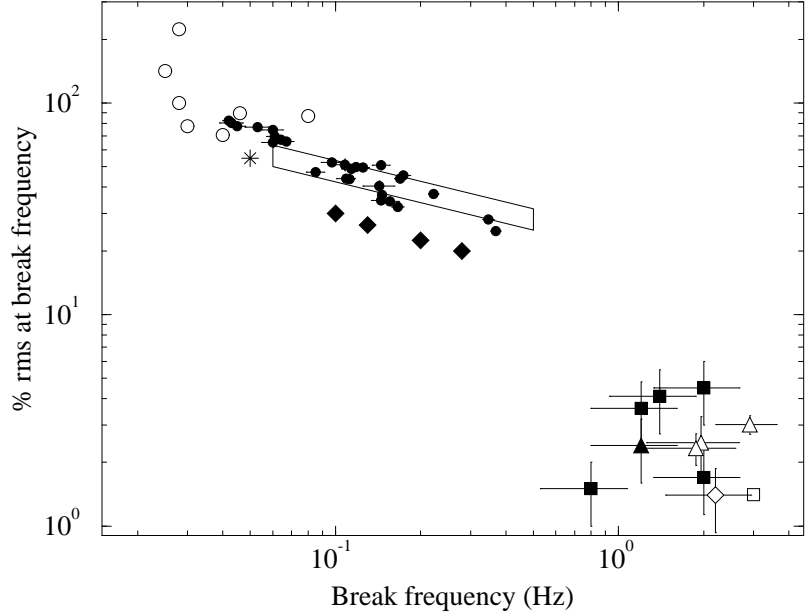}{250pt}{270}{70}{70}{-140}{400}
\caption[f3.ps]{Relation between break frequency and power
           density at the break.
           Filled circles: Cyg~X--1 in the LS (Belloni \& Hasinger 1990a);
           filled diamonds: GRO~J1719--24 (van der Hooft et al. 1996);
           asterisk: GRO~J0422+32 (Grove et al. 1994);
           open circles: GS~2023+338 in the LS (Oosterbroek 1995);
           filled squares: GX~339--4 in the VHS (Miyamoto et al. 1991, 1993);
           open diamond: GS~1124--68 in the VHS (Miyamoto et al. 1993);
           filled triangle: GX~339--4 (Belloni \& Hasinger 1990b);
           open square: GS~1124--68 in the IS (Belloni et al. 1996a);
           open triangles: GX~339--4 in the IS (this paper).
           The region marked in this figure corresponds to Cyg~X--1 LS data
           from Crary et al. 1996).}
\label{b&h}
\end{figure}

The picture that emerges from our comparison of the EXOSAT data of
GX~339--4 to data on other black hole candidates is one in which BHC's
move from the low state via an intermediate state to the high state as
the mass accretion rate increases.  On this trajectory, as a function
of $\dot M$, the power spectral break frequency increases and its rms
decreases in the way illustrated by Fig~\ref{b&h}; the disappearance
of the band-limited noise in the HS could be nothing else but the
extreme consequence of this process.  In the 40--145 keV band the flux
drops and the spectrum steepens with $\dot M$; at lower energy there
is a general increase of the ultrasoft component, but, as has been
remarked previously (Tanaka \cite{tanaka92}), not in a way that is
strictly correlated with the properties of the power law component.
We note that the properties of the peculiar flat-topped outburst of
GRO 1719-24, with an increasing break frequency (van der Hooft
\cite{vanderhooft96}) and at high energy a steepening spectrum and a
slightly decreasing flux then suggest that the mass accretion {\it
increased} during the plateau phase of this outburst.

The similarity between the IS and the VHS band limited noise power
spectra is remarkable, and it is a major challenge to explain why two
such similar states would be separated by one (the HS) with no
detectable band limited noise.  A clue might be the fact that the VHS
power spectra are violently variable, with rapid transitions between a
band limited noise and a weak power law power spectrum (Miyamoto et
al.  \cite{miyamoto91}).

{\bf Note.}  When this paper was about to be submitted, we analyzed
the recent RXTE data on Cyg~X-1 in the ``high'' state (Belloni,
Mendez, van der Klis et al.  1996, submitted to ApJ Lett.).  We found
its properties to be entirely compatible with that of the intermediate
state discussed here, which means that there are now three black hole
candidates that have shown these properties.

\acknowledgements

MK would like to thank various participants of the 1996 Aspen winter
meeting on Black Holes for pleasant and extremely fruitful
discussions.  MM wish to thank M.  Berger and T.  Oosterbroek for
useful scientific discussions.  We are very much grateful to an
anonymous referee for his comments that helped to improve the original
manuscript.  This work was supported in part by the Netherlands
Organization for Scientific Research (NWO) under grant PGS 78-277.  MM
is a fellow of the Consejo Nacional de Investigaciones
Cient\'{\i}ficas y T\'ecnicas de la Rep\'ublica Argentina.

\clearpage

\clearpage

\clearpage

\clearpage

\begin{thebibliography}{}
\bibitem[1990a]{belloni90a}
        Belloni, T., \& Hasinger, G. 1990a, \aap, 227, L33
\bibitem[1990b]{belloni90b}
        Belloni, T. \& Hasinger, G. 1990b, \aap, 230, 103
\bibitem[1996a]{belloni96a}
        Belloni, T., van der Klis, M., Lewin, W. H. G., van Paradijs,
        J., Dotani, T., Mitsuda, K., \& Miyamoto, S. 1996, \aap,
        submitted
\bibitem[1996b]{belloni96b}
        Belloni, T., M\'endez, M., van der Klis, M., Hasinger, G.,
        Lewin, W. H. G., van Paradijs, \apjl, in press
\bibitem[1994]{berger94}
        Berger, M., \& van der Klis, M. 1994, \aap, 292, 175
\bibitem[1996]{crary96}
        Crary, D. J., et al. 1996, \apjl, 462, L71
\bibitem[1992]{callanan92}
	Callanan, P. J., Charles, P. A., Honey, W. B., \& Thorstensen
	J. R. 1992, \mnras, 259, 395
\bibitem[1979]{doxsey79}
        Doxsey, R., Grindlay, J., Griffiths, R., Bradt, H.,
        Johnston, M., Leach, R., Schwartz, D., \& Schwartz, J.
        1979, \apjl, 228, L67
\bibitem[1994]{ebisawa94}
        Ebisawa, K., et al. 1994, \pasj, 46, 375
\bibitem[1991]{grebenev91}
        Grebenev, S. A., Syunayaev, R. A., Pavlinskii, M. N. \&
        Dekhanov, I. A. 1991, SvA Lett., 17, 413
\bibitem[1993]{grebenev93}
        Grebenev, S. A., et al. 1993, \aaps, 97, 281
\bibitem[1994]{grove94}
        Grove, J. E., et al. 1994 AIP Conference Proc. 304, 192
\bibitem[1992]{iga92}
        Iga, S., Miyamoto, S., \& Kitamoto, S. 1992, in Frontiers
        of X-ray Astronomy, ed. Y. Tanaka \& K. Koyama, 309
\bibitem[1986]{ilovaisky86}
        Ilovaisky, S. A., Chevalier, C., Motch, C., \&
        Chiappetti, L. 1986, \aap, 164, 67
\bibitem[1987]{imamura87}
        Imamura, J. N., Steiman-Cameron, T. Y., \& Middleditch, J.
        1987, \apjl, 314, L11
\bibitem[1990]{imamura90}
        Imamura, J. N., Kristian, J., Middleditch, J., \&
        Steiman-Cameron, T. Y. 1990, \apj, 365, 312
\bibitem[1994]{kuulkers94}
        Kuulkers. E., van der Klis, M., Oosterbroek, T., Asai, K.,
        Dotani, T., van Paradijs, J., \& Lewin, W. H. G. 1994, \aap
        289, 795
\bibitem[1984]{maejima84}
        Maejima, Y., Makishima, K., Ogawara, Y., Oda, M., Tawara, Y.,
        \& Doi, K. 1984, \apj, 285, 712
\bibitem[1986]{makishima86}
        Makishima, K., Maejima, Y., Mitsuda, K., Bradt, H. V.,
        Remillard, R. A., Tuohy, I. R, Hoshi, R., \& Nakagawa, M.
        1986, \apj, 308, 635
\bibitem[1988]{makishima88}
        Makishima, K., \& Miyamoto, S. 1988, \iaucirc, No. 4653
\bibitem[1973]{markert73}
        Markert, T. H., Canizares, C. R., Clark, G. W.,
        Lewin, W. H. G., Schnopper, H. W., \& Sprott, G. F. 1973,
        \apjl, 184, L67
\bibitem[1991]{miyamoto91}
        Miyamoto, S., Kimura, K., Kitamoto, S., Dotani, T., \&
        Ebisawa, K. 1991, \apj, 383, 784
\bibitem[1992]{miyamoto92}
        Miyamoto, S., Kitamoto, S., Iga, S., Negoro, H., \&
        Terada, K. 1992, \apjl, 391, L21
\bibitem[1993]{miyamoto93}
        Miyamoto, S., Iga, S., Kitamoto, S., \& Kamado, Y. 1993,
        \apjl, 403, L39
\bibitem[1994]{miyamoto94}
        Miyamoto, S., Kitamoto, S., Iga, S., Hayashida, K., \&
        Terada, K. 1994, \apj, 435, 398
\bibitem[1982]{motch82}
        Motch, C., Ilovaisky, S. A., \& Chevalier, C. 1982, \aap,
        109, L1
\bibitem[1983]{motch83}
        Motch, C., Ricketts, M. J., Page, C. G., Ilovaisky, S. A.,
        \& Chevalier, C. 1983, \aap, 119, 171
\bibitem[1985]{motch85}
         Motch, C., Ilovaisky, S. A., Chevalier, C., Angebault, P.
         1985, Space Sci. Rev., 40, 219
\bibitem[1996]{narayan96}
        Narayan, R. 1996, \apj, 462, 136
\bibitem[1982]{nolan82}
        Nolan, P. L., Gruber, D. E., Knight, F. K., Matteson, J. L.,
        Peterson, L. E., Levine, A. M., Lewin, W. H. G.,
        \& Primini, F. A. 1982, \apj, 262, 727
\bibitem[1972]{oda71}
        Oda, M., Gorenstein, P., Gursky, H., Kellogg, E.,
        Schreier, E., Tananbaum, H., \& Giacconi, R. 1971,
        \apjl, 166, L1
\bibitem[1976]{oda76}
        Oda, M., Doi, K., Ogawara, Y., Takagishi, K., \&
        Wada, M. 1976, \apss, 42, 223
\bibitem[1995]{tim95}
        Oosterbroek, T. 1995, Ph. D. Thesis, University of Amsterdam
\bibitem[1982]{ogawara82}
        Ogawara, Y., Mitsuda, K., Masai, K., Vallerga, J. V., Cominsky,
        L. R., Grunsfeld, J. M., Kruper, J. S., \& Ricker, G. R.,
        \nat, 295, 675
\bibitem[1995]{parmar95}
        Parmar, A. N., Lammers, U., \& Angelini, L. 1995,
        Legacy, 6, 27
\bibitem[1983]{ricketts83}
        Ricketts, M. J. 1983, \aap, 118, L3
\bibitem[1979]{samimi79}
        Samimi, J., et al. 1979, \nat, 278, 434
\bibitem[1990]{steiman90}
        Steiman-Cameron, T. Y., Imamura, J. N., Middleditch, J.,
        \& Kristian, J. 1990, \apj, 359, 197
\bibitem[1992]{tanaka92}
        Tanaka, Y. 1992, in Proc. Ginga Memorial Symp. on
        Astrophysics (Tokyo: ISAS), 19
\bibitem[1972]{tananbaum72}
        Tananbaum, H., Gursky, H., Kellogg, E., Giacconi, R.,
        \& Jones, C. 1972, \apjl, 177, L5
\bibitem[1981]{turner81}
        Turner, M. J. L., Smith, A., \& Zimmermann, M. U. 1981,
        Space Sci. Rev., 30, 513
\bibitem[1996]{vanderhooft96}
        van der Hooft, F., et al. 1996, \apjl, 458, L75
\bibitem[1989]{vanderklis89}
        van der Klis, M. 1989, in Timing Neutron Stars,
        ed. H. \"Ogelman \& E. P. J. van den Heuvel, NATO ASI
        Series C262, 27
\bibitem[1994]{vanderklis94a}
        van der Klis, M. 1994, \aap, 283, 469
\bibitem[1995]{vanderklis95}
        van der Klis, M. 1995, in Proc.~NATO~ASI Lives of the
        neutron stars, ed. M. A. Alpar et al., 301
\bibitem[1996]{vanderklis96}
        van der Klis M. 1996, paper presented at the 1996 Aspen
        Winter Meeting on Black Hole Transients
\bibitem[1988]{white88}
        White, N. E., \& Peacock, A. 1988, in X-ray Astronomy
        with EXOSAT, ed. R. Pallavicini \& N. E. White, Mem.
        Soc. Astron. Ital., 59, 7

\end{thebibliography}
\end{document}